# Experimental Investigation of Weak Non-Mesonic Decay of $_\Lambda^{10}Be$ Hypernuclei at CEBAF

S. Majewski[1], L. Majling[2], A. Margaryan[3], L. Tang[1]

[1]*TJNAF, Newport News, VA 23606, USA*
[2]*Nuclear Physics Institute, CZ-25068 Rez, Czech Republic*
[3]*Yerevan Physics Institute, 375036 Yerevan, Armenia*

Hypernuclei are convenient laboratory to study the baryon-baryon weak interaction and associated effective Hamiltonian. The strangeness changing process, in which a Λ hyperon converts to a neutron with a release up to 176 MeV, provides a clear signal for a conversion of an s-quark to a d-quark. We propose to perform a non-mesonic weak decay study of $_\Lambda^{10}Be$ hypernuclei using the (e,e'K$^+$) reaction. These investigations will fully utilize the unique parameters of the CEBAF CW electron beam and RF system and are enabled by (1) the use of new detector for alpha particles based on the recently developed RF timing technique with picosecond resolution and (2) the small angle and large acceptance kaon spectrometer-HKS in Hall C.

## Introduction

The field of weak decay of Λ-hypernuclei has experienced an impressive progress in the last few years. We mention here only the new experimental results [1], and calculations [2] and [3]. For more information we refer to [4]. The main problem concerning to the weak decay of Λ-hypernuclei is the disagreement between the theoretical and experimental values for the ratio $\Gamma_n/\Gamma_p$ between the neutron- and the proton- induced widths. The theoretical calculations underestimate the central data points for all considered hypernuclei, although the large experimental error bars do not permit any definitive conclusion. The data are quite limited and not precise since it is difficult to detect the products of the non mesonic decays, especially for the neutron-induced one. In order to solve the $\Gamma_n/\Gamma_p$ puzzle, many attempts have been made up to now. Among these we recall the introduction of mesons heavier than the pion in the ΛN transition potential; the role of two-nucleon stimulated decay; the description of the short range baryon-baryon interaction in terms of quark degrees of freedom.

## Method

The present experimental resolution for the detection of the outgoing nucleons does not allow to identify the final state of the residual nucleus in the process $_\Lambda^A Z \to {}^{A-2}Z + nn$ and $_\Lambda^A Z \to {}^{A-2}(Z-1) + np$. As a consequence, the measurements supply decay rates averaged over several nuclear final states.

We focus our attention on one peculiar case [5, 6]. It is well known that removing one nucleon from $^9Be$ or $^9B$ results $^8Be^*$ [7]:

$$^9Be \to p + (^8Li \xrightarrow{\beta^-} {}^8Be^*) ; \qquad ^9Be \to n + {}^8Be^* ;$$

$$^9B \to n + (^8B \xrightarrow{\beta^+} {}^8Be^*) ; \qquad ^9B \to p + {}^8Be^* .$$

Prevailing part of the final states of the residual nuclei ultimately decay into α-α channels (see Table 1).

TABLE 1: Spectroscopic factors extracted from the experimental results for $^9Be(p,d)^8Be$ reaction [8] are compared with predictions of various calculations based on the intermadiate coupling shell model [9].

| $^9Be$ | | | | | Calculated | | measured |
|---|---|---|---|---|---|---|---|
| $J_i^\pi$ | $T_i$ | $E_i$ | $\Gamma_i$ (keV) | decay | $E_i$ | $S_i^n$ | $S_i^n$ |
| $0^+$ | 0 | 0.00 | 6.8 eV | $\alpha$ | 0.00 | 0.550 | 0.67±0.14 |
| $2^+$ | 0 | 3.04 | 1500 | $\alpha$ | 3.09 | 0.755 | 1.49±0.23 |
| $4^+$ | 0 | 11.4 | ~3500 | $\alpha$ | 10.30 | 0.000 | no fit |
| $2^+$ | 1 (0) | 16.63 | 108 | $\alpha, \gamma$ | 16.76 | 0.505 | |
| $2^+$ | 0 (1) | 16.92 | 74 | $\alpha, \gamma$ | 16.89 | 0.430 | 1.14±0.19 |
| $1^+$ | 1 | 17.64 | 11 | p, $\gamma$ | 17.59 | 0.223 | 0.27±0.05 |

Through this unique process it would be possible to identify these final states. These events were recognized easily as "hammer tracks" in the emulsion [10].

So, due to these specific properties of the core nuclei $^9Be$ and $^9B$, it may be possible to measure the partial decay widths for the $^{10}_\Lambda Be$ and $^{10}_\Lambda B$ hypernuclei.

### Partial Decay Widths

The nonmesonic decay rate $\Gamma_{nm}$ can be written as [11]:

$$\Gamma_{nm} = \sum_\tau \Gamma^\tau = \sum_\tau \sum_i \Gamma_i^\tau ,$$

( $\tau$ = n (neutron) or p (proton)), where the partial width $\Gamma_i^\tau$ is

$$\Gamma_i^\tau = \left| \left\langle \Psi^{A-2}(\{i\}) \otimes \Psi^{NN}(JT) \middle| V_{weak} \middle| [\Psi^{A-1}(\{c\}) \otimes \Psi^\Lambda(\tfrac{1}{2})]^J \right\rangle \right|^2$$

(the shorthand notation $\{i\} \equiv E_i, J_i, T_i, \tau_i$ and $\{c\} \equiv E_c, J_c, T_c, \tau_c$ is used).

It is possible to factorize this expression as:

$$\Gamma_i^\tau = \sum_{SJ} G_J^2(\{c\},\{i\},\tau LSJ) \cdot w_{l\tau}^{SJ} ,$$

with

$$w_{l\tau}^{SJ} = \left| \sum_{LS} \left\langle l_1 l_2 : L'S'JT \middle| V_{weak} \middle| \tau s_\Lambda : L = lSJ \right\rangle \right|^2 ,$$

the spin-dependent part of the weak interaction matrix element and $G_J(\{c\},\{i\},\tau LSJ)$, the N$\Lambda$ pair fractional parentage coefficient, defined through

$$\left[\Psi^{A-1}(\{c\}) \otimes \Psi^\Lambda(\tfrac{1}{2})\right]^J \equiv \left| l^k(E_c J_c T_c) \otimes s_\Lambda : J \right\rangle =$$

$$= \sum_{i, t\!L S J} G_J(\{c\}, \{i\}, t\!L S J) \cdot \left| l^{k-1}(E_i J_i T_i) \otimes t\!l s_\Lambda (L S J) : J \right\rangle,$$

$$G_J(\{c\}, \{i\}, t\!L S J) = \sqrt{k} \cdot (T_i \tau_i \tfrac{1}{2} \tau | T_c \tau_c) \cdot \sum_j g_i^c(lj) \cdot U(J_i j J \tfrac{1}{2} : J_c J) \cdot U(l \tfrac{1}{2} J \tfrac{1}{2} : jS).$$

In Fig. 1 the relevant states of $A = 8$ isotopes are displayed. The similarity of the structure of the $\Gamma_{\alpha\alpha i}^n(_\Lambda^{10}Be)$ and $\Gamma_{\alpha\alpha i}^p(_\Lambda^{10}B)$ is clearly seen.

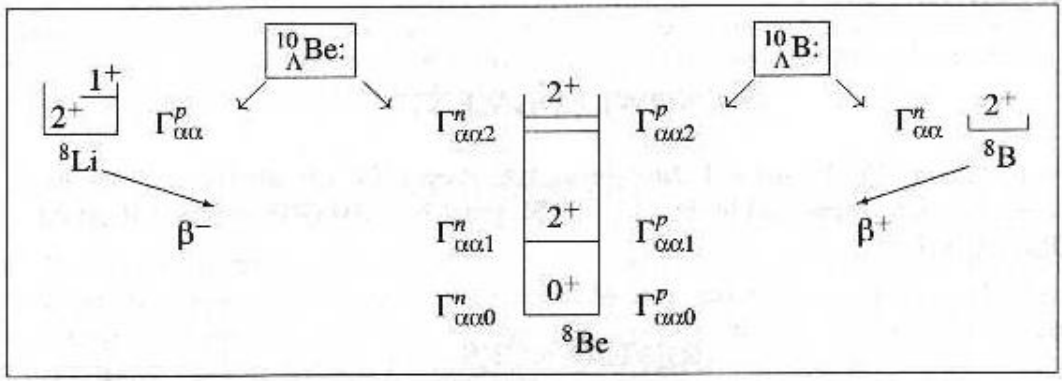

FIG. 1: Notation of the partial widths $\Gamma_{\alpha\alpha i}^\tau$

## Results

Table 2 demonstrates that these partial widths are various combinations of four matrix elements (eight for different $\tau$), hence their study offers a unique possibility to determine all needed matrix elements of the weak interaction [11] and can shed some light to the resolution of the $\Gamma^n / \Gamma^p$ puzzle [12].

TABLE 2: Partial $\Gamma_{\alpha\alpha i}^\tau(_\Lambda^{10}Be)$ and total $\Gamma_{tot}^\tau(_\Lambda^{10}Be)$ decay widths.

| p-shell proton | $(w_{1p}^{SJ})$ | s-shell proton |
|---|---|---|

$\Gamma_{\alpha\alpha}^p(_\Lambda^{10}Be) = 0.441 w_{1p}^{01} + 0.491 w_{1p}^{11} + 0.548 w_{1p}^{12} + 0.157 w_{1p}^{10}$

$\Gamma_{\alpha\alpha2}^p(_\Lambda^{10}B) = 0.096 w_{1p}^{01} + 0.520 w_{1p}^{11} + 0.439 w_{1p}^{12}$

$\Gamma_{\alpha\alpha1}^p(_\Lambda^{10}B) = 0.388 w_{1p}^{01} + 0.0511 w_{1p}^{11} + 0.408 w_{1p}^{12}$

$\Gamma_{\alpha\alpha0}^p(_\Lambda^{10}B) = 0.412 w_{1p}^{01} + 0.206 w_{1p}^{11}$

$\Gamma_{tot}^p(_\Lambda^{10}B) = 1.187 w_{1n}^{01} + 0.822 w_{1n}^{11} + 1.248 w_{1n}^{12} + 0.117 w_{1n}^{10} + 0.354 w_{0n}^{00} + 1.271 w_{0n}^{11}$

$\Gamma_{tot}^p(_\Lambda^{10}Be) = 0.569 w_{1p}^{01} + 0.535 w_{1p}^{11} + 0.981 w_{1p}^{12} + 0.165 w_{1p}^{10} + 0.437 w_{0p}^{00} + 1.313 w_{0p}^{11}$

| p-shell neutron | $(w_{1n}^{SJ})$ | s-shell neutron |
|---|---|---|

$\Gamma_{\alpha\alpha}^n(_\Lambda^{10}B) = 0.141w_{1n}^{01} + 0.489w_{1n}^{11} + 0.505w_{1n}^{12}$

$\Gamma_{\alpha\alpha2}^n(_\Lambda^{10}Be) = 0.096w_{1n}^{01} + 0.520w_{1n}^{11} + 0.439w_{1n}^{12}$

$\Gamma_{\alpha\alpha1}^n(_\Lambda^{10}Be) = 0.388w_{1n}^{01} + 0.0511w_{1n}^{11} + 0.408w_{1n}^{12}$

$\Gamma_{\alpha\alpha0}^n(_\Lambda^{10}Be) = 0.412w_{1n}^{01} + 0.206w_{1n}^{11}$

$\Gamma_{tot}^n(_\Lambda^{10}Be) = 1.187w_{1n}^{01} + 0.822w_{1n}^{11} + 1.248w_{1n}^{12} + 0.117w_{1n}^{10} + 0.354w_{0n}^{00} + 1.271w_{0n}^{11}$

$\Gamma_{tot}^n(_\Lambda^{10}B) = 0.569w_{1n}^{01} + 0.535w_{1n}^{11} + 0.981w_{1n}^{12} + 0.165w_{1n}^{10} + 0.437w_{0n}^{00} + 1.313w_{0n}^{11}$

Table 3 demonstrates the existing experimental opportunities for measuring these partial widths and $\sigma_{prod}(_\Lambda^{10}B)$, $\sigma_{prod}(_\Lambda^{10}Be)$.

TABLE 3: Sources of $\sigma_{prod}(_\Lambda^{10}B)$, $\sigma_{prod}(_\Lambda^{10}Be)$, $\Gamma_{\alpha\alpha i}^\tau(_\Lambda^{10}Be)$ and $\Gamma_{\alpha\alpha i}^\tau(_\Lambda^{10}B)$

| $_\Lambda^{10}B$ | | | | $_\Lambda^{10}Be$ | | | |
|---|---|---|---|---|---|---|---|
| $\Gamma_{\alpha\alpha2}^p$ | em | c2 | DN | $\Gamma_{\alpha\alpha2}^n$ | em | c1 | DN |
| $\Gamma_{\alpha\alpha1}^p$ | em | | DN | $\Gamma_{\alpha\alpha1}^n$ | em | | DN |
| $\Gamma_{\alpha\alpha0}^p$ | | | | $\Gamma_{\alpha\alpha0}^n$ | | | |
| $\Gamma_{\alpha\alpha}^n$ | em | | | $\Gamma_{\alpha\alpha}^p$ | em | | |
| $\Gamma_{tot}^p$ | F1 | | | $\Gamma_{tot}^n$ | f2 | | |
| $\Gamma_{tot}^n$ | F1 | | | $\Gamma_{tot}^p$ | f2 | | |
| $\sigma_{prod}$ | F1 | c2 | | $\sigma_{prod}$ | f2 | c1 | |

| Notation | |
|---|---|
| F1: $^{10}B(K^-,\pi^-)_\Lambda^{10}B$ | f2: $^{10}B(K^-,\pi^0)_\Lambda^{10}Be$ |
| c2: $^{10}B(e,e'K^0)_\Lambda^{10}B$ | C1: $^{10}B(e,e'K^+)_\Lambda^{10}Be$ |
| em: emulsion | DN: Rel. Ion Coll. |

**Time-Zero Nuclear Fragment Detector for Hypernuclear Studies at CEBAF**

The time precision limit for the current systems consisting of particle detectors, timing discriminators and time to digital converters is about 100 ps (FWHM). We report here a new picosecond timing concept for heavily ionizing particles based on the RF analysis of produced and accelerated secondary electrons [13]. The operational principles of the new timing technique for fission fragments are presented in Fig. 2.

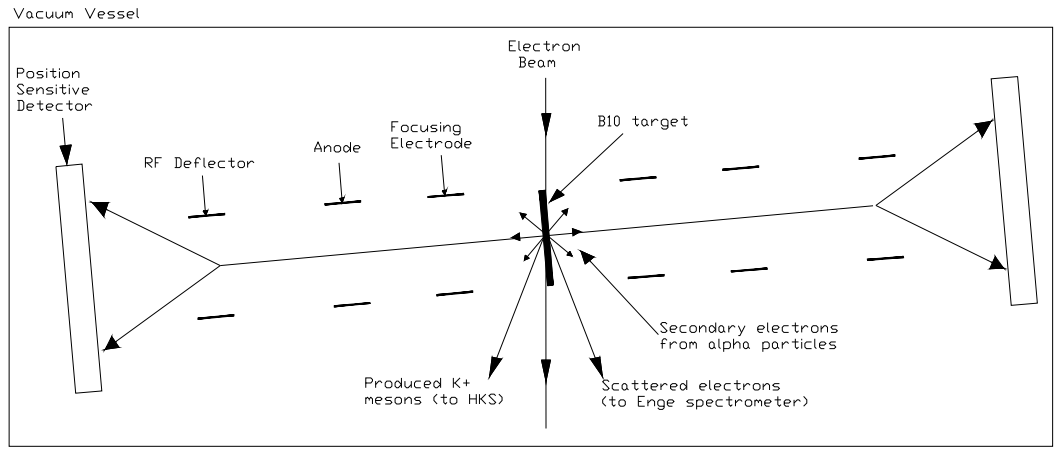

FIG. 2: The schematic diagram of the RF timing system based on secondary electron emission.

Nuclear fragments passing through target surface produce secondary electrons in proportion with their energy loss in matter(~4secondary electrons for 5 MeV α particles and about 150 for fission fragments). The secondary electrons are emitted with a large angular spread, but with energies mainly on the order of a few electron volts. The secondary electrons are accelerated up to several keV, focused, and transported to the RF resonator deflector. Passing through the RF deflector, the secondary electrons change their direction as a function of the relative phase of electromagnetic field oscillations in the RF cavity, thereby transforming time information into deflection angle. Coordinates of the secondary electrons are measured with the help of a position sensitive detector.

In this project, we propose to use the oscilloscopic method of RF timing, which result in a circular pattern on the detector plane. We will use recently developed dedicated RF deflecting system for non-relativistic secondary electrons at a frequency of f = 500 MHz . The circular scan of a 2.5 keV electron beam on the phosphor screen is demonstrated in the Fig. 3.

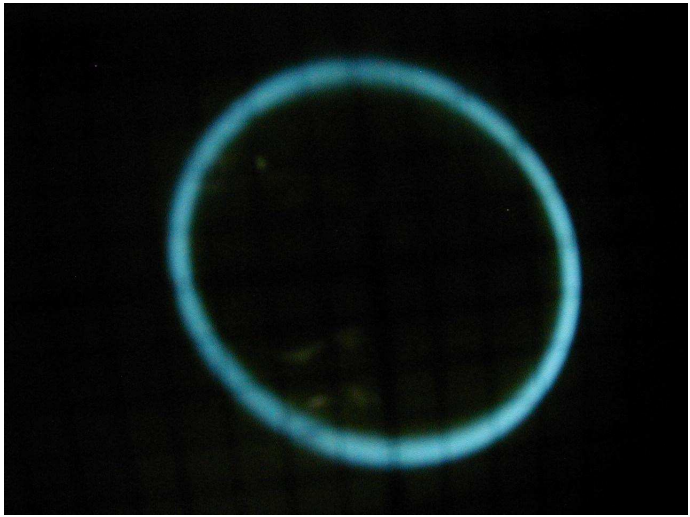

FIG. 3: Image of circularly scanned 2.5 keV energy electron beam on the phosphor screen.

The time resolution of the system presented in Fig.2 consists of several factors [14, 15]:

**1. Physical Time Resolution of Electron Tube:** The minimal physical time dispersion of the electron optics is determined by chromatic aberration due to the electron initial energy spread- $\Delta\varepsilon$. The time spread $\Delta\tau_t$ due to this effect in the case of a uniform accelerating electric field near the photocathode plane- E in V/cm is $\Delta\tau_t = 2.34 \times 10^{-8} (\Delta\varepsilon)^{1/2}$/E sec. For $\Delta\varepsilon = 4$ eV and E = 10 kV/cm we obtain $\Delta\tau_t \approx 4$ ps.

**2. Time Dispersion due to Coulomb Repulsion:** The secondary electron trajectory displacement due to Coulomb repulsion effects is one of the main factors limiting timing resolution of the electron tube in the sub-picosecond region. The influence of the Coulomb repulsion effects on the dynamics of electron bunches in electron tubes has been studied numerically [16]. It has been evaluated that time dispersion due to Coulomb repulsion - $\Delta\tau_q \sim 10^{-14} \times n$ sec, where n is the number of electrons in a single event. For $\alpha$-particles $\Delta\tau_q < 10^{-13}$ sec, but for fission fragments $\Delta\tau_q \sim 2 \times 10^{-12}$ sec.

**3. Technical Time Resolution of Electron Tube:** Electron tube is a device with precise electron focusing in the effective electron transit time equalization. The time precision limit for the whole system consisting of photo-cathode, accelerating, focusing and transporting parts in the carefully designed system is estimated to be on the order of 0.1 ps (FWHM) for point like photocathode [17].

**4. Technical Time Resolution of the RF Deflector:** By definition, the technical time resolution is: $\Delta\tau_d = d/v$, where $d$ is the size of the electron beam spot or the position resolution of the secondary electron detector if the electron beam spot is smaller, while $v$ is the scanning speed: $v = 2\pi R/T$, here T is the period of rotation of the field, R is the radius of the circular sweep on the position sensitive detector. For example, if $T = 2 \times 10^{-9}$ sec (f = 500 MHz), R = 2 cm, and d = 0.5 mm, we have $v \geq 0.5 \times 10^{10}$ cm/sec and $\Delta\tau_d \leq 10 \times 10^{-12}$ sec.

## Position Sensitive Detector

Position sensitive detectors for the RF timing technique should have the following properties:

- High detection efficiency for 2.5-10 keV single electrons;
- Position sensitivity, for example, $\sigma_x \leq$ of 0.5 mm;
- High rate capability.

These three requirements are to a different degree realized by several types of detectors. They include vacuum- based devices such as the Multi-anode PMT [18, 19], Micro Channel Plates (MCP) [20] as well as an array of Si PIN [21] and Avalanche Photodiodes (APD) [22] or APD working in a Geiger mode (SiPM) [23].

However, the best approach is the development of a dedicated multi-anode PMT with circular anode structure.

Position determination can be performed in two basic architectures:

1) Direct readout: array of small pixels, with one readout channel per pixel, such as available with avalanche Si diodes. Position resolution in this case is about or better than the size of readout cell.

2) Interpolating readout: position sensor is designed in such a way that measurement of several signals (amplitudes or/and times) on neighboring electrodes defines event position. Position resolution limit for both cases is $\Delta x/x \sim 10^{-3}$ [24].

After analysis of the available technical solutions for position sensors, the best solution in our opinion is the development of a dedicated position sensitive technique based on the MCP or/and APD with circular anode structure and interpolating readout schemes for position determination of secondary electrons with relative precision close to $10^{-3}$. It is worth to mention that sensors based on Si diodes can be operated near high current (~3 mA) electron beam [25].

From the temporal resolution budget discussed above one can conclude that the expected parameters of the technique are:

a) Internal time resolution for each photo-electron of about 4 ps;

b) Technical time resolution in the range of $(2-10)\times10^{-12}$ sec for f = 500 MHz RF deflector;

c) Absolute calibration of the system better than $10^{-12}$ sec is possible, which is an order of magnitude better than can be provided by a regular timing technique.

## Nonmesonic Weak Decay Study of $^{10}_{\Lambda}Be$ at CEBAF

A pioneering experiment in Lambda hypernuclear spectroscopy, undertaken at the Thomas Jefferson National Accelerator Facility (Jlab) [26] demonstrates that the $(e, e^{'}K^{+})$ reaction can be effectively used as a high resolution tool to study hypernuclear spectra and its use should be vigorously pursued [27, 28]. The experiment used the high-precision, continuous electron beam at Jlab, and a special arrangement of spectrometer magnets to measure the spectrum from nuclear targets using the $(e, e^{'}K^{+})$ reaction. We propose to use the same experimental setup and reaction $^{10}B(e, e^{'}K^{+})^{10}_{\Lambda}Be$ to produce and investigate the weak nonmesonic decay of $^{10}_{\Lambda}Be$ hypernuclei. Neutron induced weak nonmesonic decay of $^{10}_{\Lambda}Be$ hypernuclei will result two back-to-back $\alpha$ particles from three energetic levels of residual $^{8}Be^{*}$ nucleus (see Fig. 1). The exit of low energy $\alpha$-particles from $^{10}B$ target can be suppressed by proper selection of the target thickness. For example from ~3 mg/cm$^2$ thickness of $^{10}B$ target effectively exit only two $\alpha$-particles from ~16 MeV levels of residual $^{8}Be^{*}$ nucleus. In such a way properly selected target thickness act as a filter for $^{8}Be^{*}$ levels. Meanwhile due to fine time microstructure of Jlab electron beam (2 ps bunches each 2 ns interval), the Jlab is the best place where the new timing technique for nuclear fragments can be applied.

We propose to carry out this measurement in Hall C, utilizing the HKS spectrometer system for the detection of the produced $K^+$ and scattered electron in coincidence with delayed signals from two $\alpha$-particles. By this way the bound regions will be fixed and $^{10}_{\Lambda}Be$ hypernuclei production cross section will be determined as well. The experimental layout is shown in Fig. 4. Electron beam passing through 3 mg/cm$^2$ tilted target (the effective thickness of the target is about 10 times higher) will produce $K^+$ mesons on the bound protons due to the $\gamma^v p \rightarrow K^+ \Lambda$ reaction. The expected yield of the hypernuclear states are evaluated based on the E89-009 result for $^{12}_{\Lambda}Be$ ground state in the $^{12}C(e, e\,K^+)^{12}_{\Lambda}B$ reaction [27]. The cross sections of the hypernuclear states for the target $^{10}B$ have been calculated by Motoba et al. [29]. It is noted, however, the calculated cross section vary by a factor of 2-5 depending on the choice of model parameters for the elementary reaction, hypernuclear potentials and configuration of the states etc. In the present yield estimate 100 nb/Sr cross section have been used. For the beam current of $30\,\mu$A and the target thickness of 100 mg/cm$^2$, a typical hypernuclear ($^{12}C$ target) event rate is 48.4/(100nb/Sr)/h [26]. Taking this into account, the expected yield of the hypernuclear states for the $^{10}B$ target thickness of 30 mg/cm$^2$ and the beam current 100 $\mu$A will be about 50/h. About 30% (see TABLE 1) of the weak nonmesonic decays of $^{10}_{\Lambda}Be$ hypernuclei will result two $\alpha$-particles from the ~16 MeV excited state of $^8Be^*$ residual nucleus. Each $\alpha$-particle will have ~8 MeV kinetic energy and will exit $^{10}B$ target and detected in the nuclear fragment detector. The expected exit and detection probability is about 90%. Therefore the full event ($e^-, K^+$ + two delayed $\alpha$-particles) rate will be about 10/h.

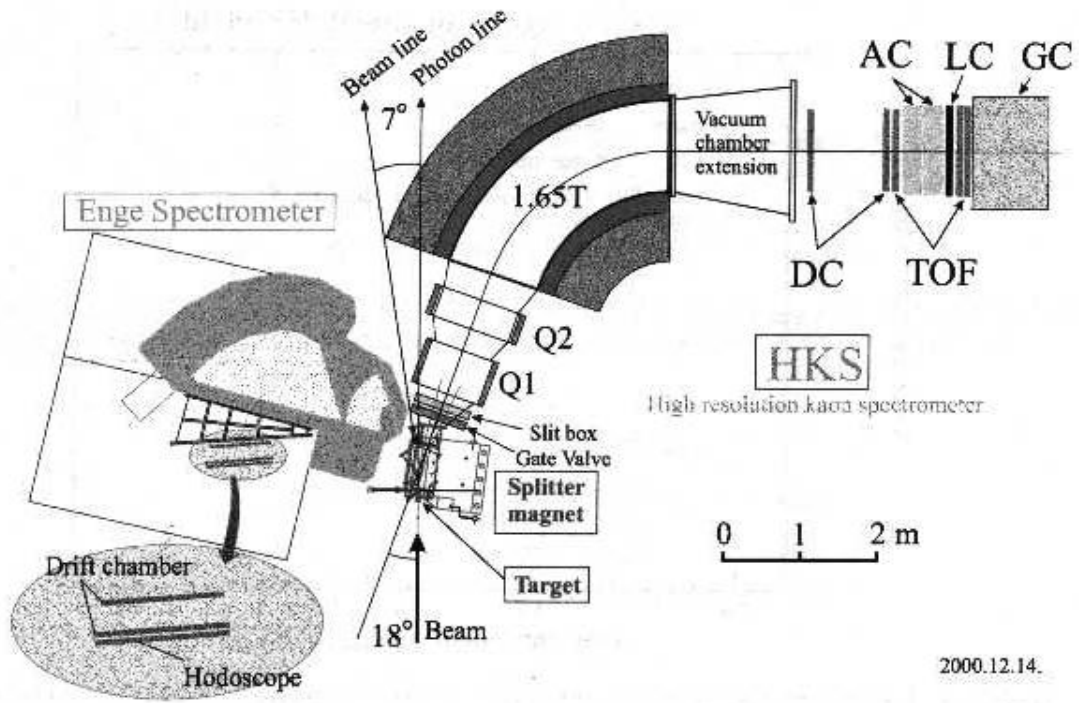

FIG. 4: The HKS general configuration with detectors. In our case $^{10}B$ target with RF timing system will be mounted in the target place of HKS.

## Summary

The non-mesonic weak decays of $\Lambda$ hypernuclei and an explanation of the $\Gamma^n/\Gamma^p$ ratio in particular are still one of the hot topics in the hypernuclear physics [30]. In the course of searching for "new strategy" [31] we continue the discussion of suggestion presented at HYP00 conference [5]: to study non-mesonic weak decays of the $_\Lambda^{10}Be$ hypernucleus.

We propose to use the unique feature of the $^9Be$ nucleus: after removing a neutron from its ground state several groups of $\alpha$- particles appear from different excited states of the residual nucleus $^8Be^*$. Due to its salient cluster structure – $\alpha\alpha$ n$\Lambda$ it is possible to measure in the $_\Lambda^{10}Be$ hypernucleus the partial "$\alpha$ – decay widths" $\Gamma_{\alpha\alpha i}^\tau$, corresponding to states of the residual nucleus $^8Be^*(E_i, J_i^\pi, T_i)$ which decays through the $\alpha\alpha$ – channel.

In such a way we can determine one-nucleon simulated process $\Lambda n \rightarrow nn$ unambiguously. Obviously, the role of the two-nucleon stimulated process $\Lambda np \rightarrow nnp$, could be seen by detection of the $\alpha$ – particles in the decay of the hypernucleus $_\Lambda^{11}B \rightarrow {}^8Be + nnp$. The partial widths $\Gamma_{\alpha\alpha i}^\tau$ can be determined through detection of tagged $\alpha$ – particles. Such tagged $\alpha$ – particles were recognized as "hammer tracks" in the emulsion and were efficiency used for identification of $_\Lambda^8Li \rightarrow \pi^- Be^*$ [10].

Here we propose to use the HKS in Hall C, Jlab and reaction $^{10}B(e,e^!K^+)_\Lambda^{10}Be$ to produce and investigate the weak nonmesonic decay of $_\Lambda^{10}Be$ hypernuclei. In a first step we propose to use time-zero nuclear fragment detector with picosecond resolution to detect delayed $\alpha$– particles in coincidence with scattered electron and produced $K^+$ meson and determine the $\Gamma^n_{\alpha\alpha 2}(_\Lambda^{10}Be)$ together with $\sigma_{prod}(_\Lambda^{10}Be)$. Next we plan to extend the program and carry out the measurement of the other values of $\Gamma_{\alpha\alpha i}^\tau$ by using recoil distance technique.

This work is supported in part by International Science and Technology Center- ISTC, Project A-372.